\documentclass[runningheads]{llncs}
\usepackage{graphicx}
\begin{document}
\title{Towards Continuous Safety Assessment in Context of DevOps}

%
%
\author{Marc Zeller\orcidID{0000-0002-6738-7903}}

\authorrunning{M. Zeller}
%
\institute{Siemens AG, Otto-Hahn-Ring 6, 81739 Munich, Germany \\
\email{marc.zeller@siemens.com}}
\maketitle              
\begin{abstract}
Traditionally, promoted by the internet companies, continuous delivery is more and more appealing to industries which develop systems with safety-critical functions.
Since safety-critical systems must meet regulatory requirements and require specific safety assessment processes in addition to the normal development steps, enabling continuous delivery of software in safety-critical systems requires the automation of the safety assessment process in the delivery pipeline.
In this paper, we outline a continuous delivery pipeline for realizing continuous safety assessment in software-intensive safety-critical systems based on model-based safety assessment methods.

\keywords{safety assessment \and agile \and DevOps \and continuous delivery.}
\end{abstract}
\section{Introduction}
\label{sec:intro}
DevOps mixes the development and operations phases of a software product by promoting high frequency software releases which enable continuous innovation based on feedback from operations. \cite{loukides2012devops}.
DevOps uses continuous integration and test automation to build a pipeline from development to test and then to production (so-called \emph{continuous delivery}).
While companies are already implementing agile practices and continuous delivery in "non-critical" software development, safety-critical software is nowadays still developed using classical waterfall or V-model-based development processes.
Safety-critical systems must meet regulatory requirements and shall comply to safety standards (such as IEC 61508). This requires (re-)certification processes for safety compliance after each change of the system.
Thereby, introducing continuous delivery of software in this context requires to solve specific issues \cite{Zeller2020a}.
However, also in the area of safety-critical systems, the need for accelerating the delivery of software is essential to reduce the time-to-market for new features and to respond faster to changing customer/market demands or technical concerns like deploying security patches.
Hence, there is an increasing need to build a \emph{"continuous safety assessment machine"} which enables continuous assessment and delivery also for software in safety-critical systems. 

This paper provides an overview of the challenges to enable continuous delivery of software in the context safety-critical systems. We outline how model-based approaches can cope with these challenges. Moreover, we illustrate a pipeline which allows continuous safety assessment based on these approaches as a first vital step towards continuous delivery in software-intensive safety-critical systems.

The remainder of this paper is organized as follows: In Sec. \ref{sec:relatedwork} we summarize related work. Afterwards, we outline the challenges that architects and developers face in the context of continuous delivery for safety-relevant systems. In Sec.~\ref{sec:apporach}, we then sketch model-based approaches to cope with these challenges. Based on these concepts, we illustrate a pipeline to enable continuous safety assessment in Sec.~\ref{sec:safetassessment}. Afterwards, we discuss additional steps towards DevOps for safety-critical systems (Sec.~\ref{sec:SafeDevOps}). The paper is concluded in Sec.~\ref{sec:summary}

\section{Related Work}
\label{sec:relatedwork}
R-Scrum \cite{6606635} and SafeScrum \cite{hanssen2018safescrum} are existing approaches to develop safety-critical systems using agile methods, but do not show how to build a continuous delivery pipeline. Also \cite{steghoefer2019} only presents challenges how to enable agile development of safety-critical systems in large organizations.
All of these papers name traceability and continues safety validation or compliance, which is an integrated part of all sprints, as a necessary step to realize agile development in regulated environments.
First ideas to realize a continuous delivery pipeline for safety-critical software are outlined in \cite{7965323} and \cite{8990195}.
This delivery pipeline includes iterative safety analysis approaches as well as automated safety test generation execution integrated in the agile software engineering process. However, the ideas w.r.t.~methods and tools to realize the delivery pipeline are very abstract and immature.
The authors in \cite{8990195} propose a combination of component-based design, the use of contracts, modular assurance cases, and agile practices to realize continuous delivery in the development of safety-critical systems.
In contrast to existing work, we address all phases of the safety engineering life-cycle and sketch a concrete delivery pipeline for continuous safety assessment based on existing model-based safety analysis approaches and model-connected assurance cases.

\section{Challenges in Enabling Continuous Delivery for Safety-critical Systems}
\label{sec:challenges}
DevOps is an approach that shortens the gap between software development and software operation in the production environment.
Continuous delivery of software is based on flexible and scalable product definitions that focus on feature-wise clustered components which are loosely coupled and can be upgraded independently (e.g.~using microservice architectures and patterns). It uses continuous integration and test automation to build pipeline from development to test and then to production (so-called \emph{continuous delivery pipeline}).

\begin{figure}[htb]
\centering
\includegraphics[width=0.8\columnwidth]{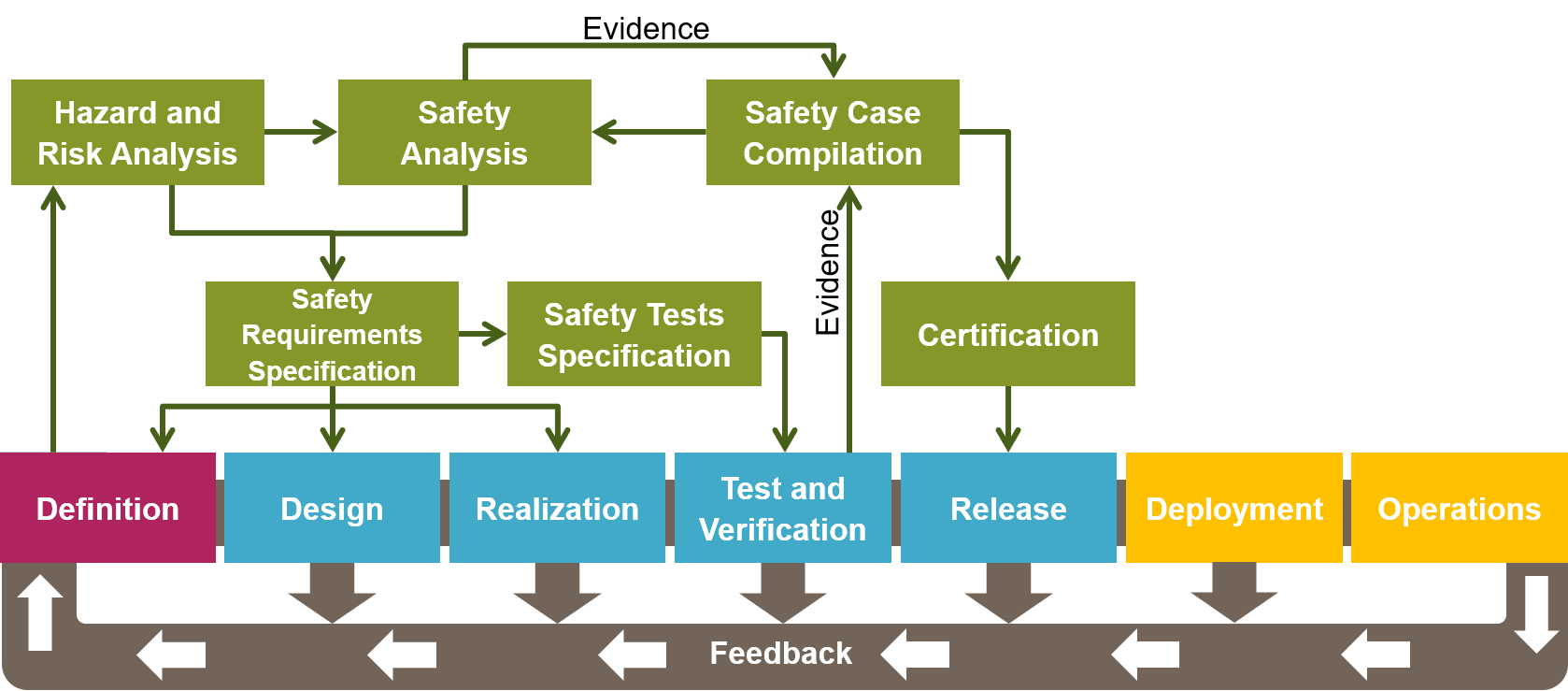}
\caption{Continuous delivery pipeline with additions for safety-critical systems development}
\label{fig:devsafeops}
\end{figure}

While continuous monitoring and measurement (often called supervision in the safety domain) is a standard design feature of safety-critical systems, continuous delivery is challenging in safety-related applications with strict regulatory requirements and safety guidelines. Since safety is a system-level property, the continuous delivery process must be lifted from software to system level and people from different engineering disciplines must be included in the continuous delivery pipeline.
Moreover, additional development steps are required in safety-critical systems, such as the HARA, the safety analysis of the architecture, the safety augmentation, and the certification (see Fig.~\ref{fig:devsafeops}). These steps are required by any safety standards (e.g.~IEC61508) and must be performed within the delivery pipeline.

Thereby, the following challenges related to the safety engineering activities need to be addressed in order to realize a delivery pipeline:

\paragraph{Hazard Analysis \& Risk Assessment:}
Today, the HARA is a manual process which requires the assessment of potential system hazards typically documented in spreadsheets. To enable continuous delivery for safety-critical systems, we need to speed-up the HARA process. This means, that we need automation for both the identification of hazards and the assessment of the risk associated to hazards. Moreover, it must be possible to identify new hazards and to adapt the risk associated with known hazards of the system.
\paragraph{Safety Analysis:}
The safety analysis techniques used today in industrial practice, such as \emph{Fault Tree Analysis (FTA)} and \emph{Failure Mode and Effect Analysis (FMEA)}, are performed manually by experts on system level. To enable continuous delivery, we need to speed-up the safety analysis process by increasing the level of automation. The changes of the software and their influence on the system safety must be reflected in the analysis automatically. Failure modes specific to highly integrated software-intensive systems such as failures due to feature interactions, emerging features, or not-wanted interactions must be integrated in the safety analysis. Moreover, the safety analyses must be linked to the system (or software/hardware) design so that changes to the system (or software/hardware) architecture can be synchronized with the safety analysis models.
\paragraph{Safety Tests:}
Since the predictable behavior of a system in the presence of faults is crucial to its safe operation, testing of safety-critical systems addresses questions such as: Are fault reactions correctly and effectively implemented? Is the timing of these reactions sufficient? Consequently, safety-critical systems require the testing of safety mechanisms that consider the fault models of the system components. Typically, tasks related to the testing of safety mechanisms are manually done by experts. In the context of a continuous delivery pipeline, the synthesis of tests for the specified failure mitigation mechanisms must be automated. Moreover, we need the possibility to inject faults into the productive system under test in virtual/real production environment without side-effects and to automatically execute the generated tests.
\paragraph{Safety Argumentation:}
Today, a safety case that describes the argumentation and references all the work products created during the safety life-cycle is manually captured in documents. Often, these work products are spread across various tools. The disparate information sources result in high accidental complexity and keeping the artifacts consistent is time-intensive and error-prone. Checking that the safety argumentation is complete and consistent with the configuration of the system, which is planned to be released, is expensive and mostly a manual process done through reviews. In order to build a continuous delivery pipeline for safety-critical systems, the creation and maintenance of the safety argumentation must be automated to reduce the time and costs involved in evolving the safety arguments.
Therefore, detailed traceability between safety arguments and evidences created during the development must be provided.
\paragraph{Orchestration of Different Disciplines:}
Since safety is a system-level property, the assessment of safety-critical software in terms of safety must be conducted on system level. Therefore, not only software engineers must be involved in the continuous delivery process, but also experts from other engineering disciplines as well as safety experts. Moreover, external assessors must be incorporated into the agile development process. In this way, at the end of each sprint/iteration, not only correct software but also the necessary assets and documentation for the independent safety assessment and certification can be delivered.

\section{Methods to Enable Continuous Delivery for Safety-critical Systems}
\label{sec:apporach}
Automation of the safety assessment process of a safety-critical software-intensive system is a viable first and mandatory step to realize DevOps in a safety-relevant context. To automate the safety assessment process, we need to automate the activities of the safety engineering life-cycle.

Model-based safety engineering and assurance approaches can help to cope with the challenges described in Section \ref{sec:challenges}. Similar to model-based engineering, model-based safety assurance \cite{Joshi2005} provides the foundation to automate tasks related to the safety engineering life-cycle. Artifacts of the safety engineering life-cycle are represented by models. With today's document-driven safety engineering practices, establishing automation is hardly possible, because information is documented in natural language text and images. Documents do not provide fine-granular traceability of information created during the system and the safety engineering life-cycle.
In contrast, models provide the possibility to capture information in a machine-readable way and enable the interlinking of different kinds of elements to establish traceability. By capturing the results of different activities performed during the safety engineering life-cycle, in the form of models, their interdependencies can be captured, and consistency and traceability can be established. Moreover, the safety engineering models can be integrated with system design artifacts provided by model-based system engineering. As a consequence, safety cases are represented as model artifacts that reference other fine-granular model elements created during the system development and safety engineering life-cycle \cite{Ratiu2015}. Based on the model-based approach, it is possible to provide methods to automate activities within the safety engineering life-cycle, e.g.~when updates of the safety-relevant system are performed.

Potential approaches to solve the challenges described in Sec.~\ref{sec:challenges} by automating the respective tasks using model-based techniques are outlined in the following:

\subsubsection{Dynamic Hazard and Risk Analysis:}
Detailed knowledge about the safety-relevant system, its environment, and the interfaces between the system and its environment is required to address the aforementioned challenges. For instance, in a robot-based industrial automation system, we need to know the robot's trajectories, the complete environment of the robot within a plant, and the tasks that the robot can perform in order to judge which hazards may occur in such a system and how to assign the associated risks to each of the identified hazards.

A possible solution is to simulate the system including its different sub-systems (software, electronics, mechanics, etc.) to identify the effects of component failures on the behavior of the system. However, in order to determine if a failure results in a hazard, the context (or the environment) in which the system operates (e.g.~the driver of a car, roads, traffic, etc.) must also be simulated. Hence, the behavior of a system at its interfaces can be observed and the consequences of a system's action on the environment can be identified and assessed in terms of occurrence and severity.
Therefore, a sophisticated simulation of the socio-technical system which consists of co-simulation of various kinds of simulations (1D, 3D, physics, human behavior, etc.) is necessary. By simulating the technical system and its environment and by including the effects of failures on humans, hazards can be identified, and the risks of the hazards can be assessed.
An example from the manufacturing domain to dynamically perform risk assessment based on a 3D simulation of the system is presented in \cite{MoncadaSPLMRZKS21}.

\subsubsection{Automated Safety Analysis:}
With \emph{Component Fault Trees (CFTs)}, there is a model- and component-based methodology for fault tree analysis \cite{Kaiser2003,KaiHofig2018,Kaiser2018}.
In CFTs, every system component is represented by a CFT element. Each element has its own in-ports and out-ports that are used to express propagation of failure modes through the tree. Similar to classic fault trees, the internal failure behavior that influences the output failure modes is modeled by Boolean gates. 

A library, which contains CFT elements for all system components, supports reusability by allowing stakeholders to create different CFTs by changing the assembly of the CFT elements according to the system architecture.
Based on the methods described in \cite{Moehrle2016,Moehrle2017}, it is possible automate the composition of CFTs.
Hence, by automatically generating mappings between the input and output failure modes, system-wide safety analysis models can be automatically created. Moreover, the safety analysis can be automatically adapted when making modifications to a system's architecture. Also analysis models for software can be automatically generated \cite{Kaukewitsch2020,Zeller2017a,Kaiser2018}.

\subsubsection{Automate Side-effect Free Fault Injection Tests:}
To ensure that the safety mitigation mechanisms are correctly implemented and executed on the target hardware, by performing side-effect free fault injection tests in a system.

Therefore, it is possible to combine the model-based safety analysis and the CFTs with the concept of \emph{data probes} \cite{froehlich16}. CFTs represent the failure behavior of a system and is the basis for the generation of system-level test cases for failure mitigation mechanisms. Data probes are special purpose library components that monitor, trace, check, and optionally, manipulate one or more variables in one or more functional components of one or more process components. Data probes can be programmed. A probe program selects probe points and tells a data probe what to do with the selected probe points in accordance with an analysis goal. Thus, data probes enable non-intrusive monitoring and side-effect free manipulation of data during the operation of software-intensive systems

Based on the failure behavior descriptions, using CFTs for model-based safety analyses allows to automatically generate test data at system-level \cite{Zeller2015a}. The resulting test cases show that the actual implementation of a system is compliant to its defined failure behavior and can be used to test the specified failure mitigation mechanisms.
Furthermore, with test case generation based on CFTs (with the data probes concept), it is possible to perform non-intrusive, side-effect free fault-injection tests and make reliable statements about the behavior of safety-relevant systems in the presence of software faults and component failures \cite{froehlich16}. These tests are derived automatically from the system’s failure behavior for the detection and handling of value and time errors. Thereby, the derived test cases provide the input, which is used to manipulate data using the probes at the input of a component and to configure monitors via the probes to observe the output of a component that is being tested.

\subsubsection{Model-based Safety Argumentation:}
A fundamental problem in today's safety engineering processes is that safety argument models are not formally integrated with the evidence models supporting the claims of the argumentation. This lack of integration hampers the effective automation of the safety assessment \cite{Ratiu2015}. Concrete examples of evidence models include hazard and safety analysis models as well as the dependability process execution documentation. These artifacts refer to the same system and they are interrelated with each other. In order to solve these challenges, the artifacts created during the safety engineering life-cycle as well as their relationships should be a part of the system's model-based reflection. In this context, the concept of the \emph{Digital Dependability Identity (DDI)} \cite{Schneider2015} can be used to capture the various artifacts of the safety life-cycle in a model-based way and to establish a relationship between the argumentation and the supporting evidence models. By establishing traceability across the artifacts, DDIs represent an integrated set of safety data models ("What is the evidence data?") that are generated by engineers and are reasoned upon in safety arguments ("How is the evidence data supporting the claim?").
A DDI provides traceability between a safety argumentation captured in form of  the \emph{Structured Assurance Case Metamodel (SACM)} \cite{sacm} and safety-related evidence models, namely, hazard and risk analysis, functional architecture, safety analysis, and safety design concept. SACM provides the assurance case backbone for creating the required traceability.
The DDI meta-model formalizing the traceability and evidence semantics is the so-called \emph{Open Dependability Exchange (ODE)} meta-model\footnote{https://github.com/Digital-Dependability-Identities and https://github.com/DEIS-Project-EU}.

\section{Continuous Safety Assessment}
\label{sec:safetassessment}
\begin{figure}[htb]
\centering
\includegraphics[width=0.85\columnwidth]{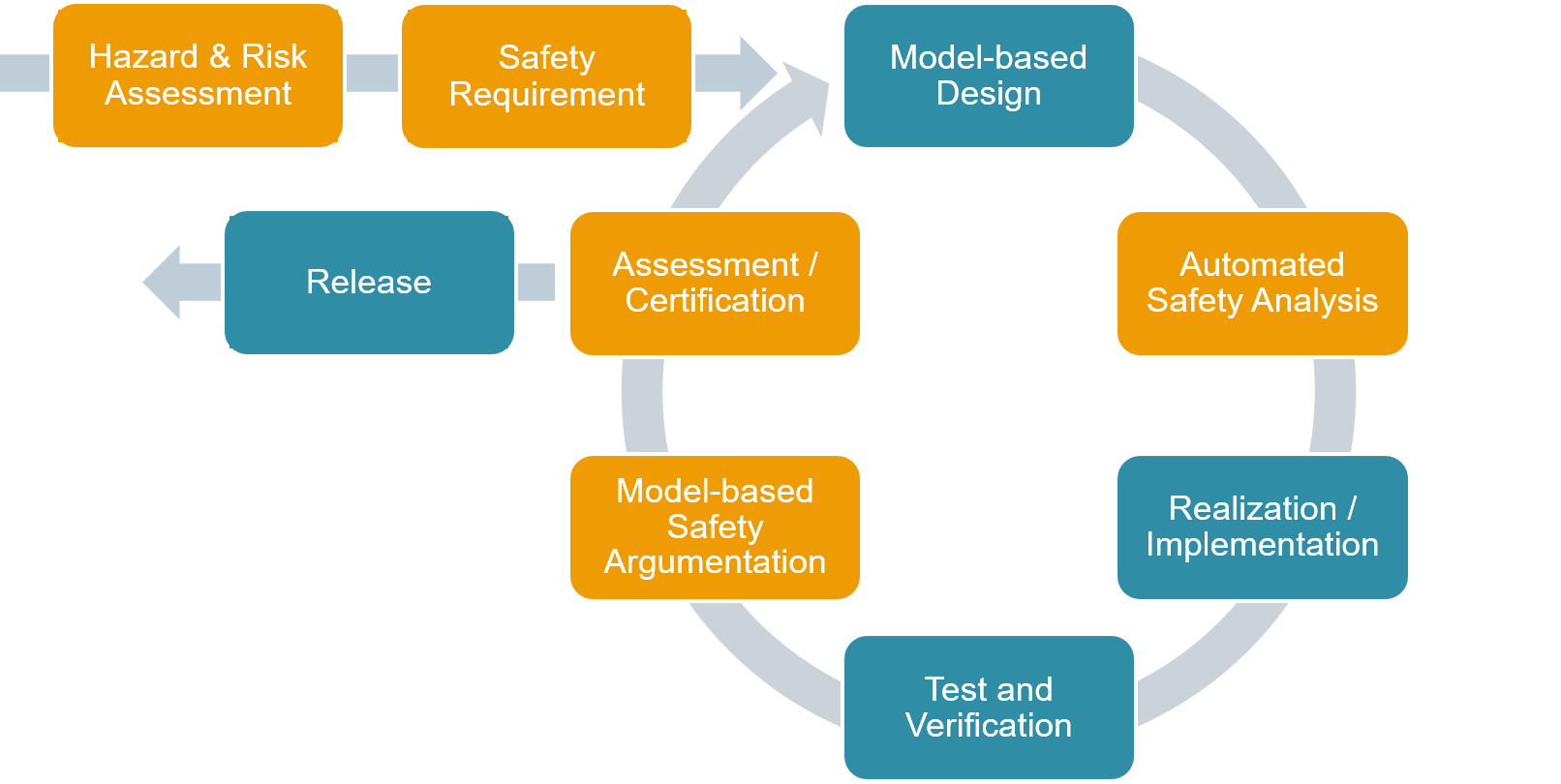}
\caption{Continuous safety assessment pipeline for safety-critical systems}
\label{fig:safetyAssessmentPipeline}
\end{figure}

In this section, we integrate the model-based approaches presented in Section~\ref{sec:apporach} to enable continuous safety assessment in a delivery pipeline as depicted in Fig.~\ref{fig:safetyAssessmentPipeline}.

\paragraph{Prerequisites:}
We assume that the Hazard \& Risk Assessment for the system has been conducted (manually). Thus, potential hazards have been identified and the associated risk was assessed. Moreover, safety goals have been derived from these hazards and the safety goals have been refined into a set of safety requirements. The safety requirements are the input for the continuous safety assessment pipeline.

\paragraph{Model-based Design:}
The system architecture is defined using a \emph{Model-Based System Engineering (MBSE)} approach (such as SysML). This includes a structural description of the system architecture including the hardware and the software architecture as well as their interactions. The safety requirements are allocated to these system elements. Moreover, the model-based design comprises of a behavioral description of the software elements, for instance in the form of UML Statecharts, Continuous Functions Chars (CFCs) or Simulink models. 

\paragraph{Automated Safety Analysis:}
Based on the model-based system description, safety analysis models in the form of CFT elements can be derived automatically for the software components when these components are defined as UML Statecharts \cite{Kaukewitsch2020}, CFCs \cite{Zeller2017a} or Simulink models \cite{Kaiser2018}. For the hardware components, the CFT elements can be composed automatically, if a library with CFT element for all necessary parts (e.g.~resistors, transistors, contactors) is predefined. Hence, a circuit diagram of a safety-critical system can be transformed automatically in a CFT model by assembling predefined CFT elements according to a given model of the circuit diagram. The system-level CFT can be built according to \cite{Hoefig2015a} based on the CFT models derived for the software, the CFT model of the different hardware parts, and the allocation of software to hardware components. Hence, the safety analysis for a system design can be conducted in an automated way. Moreover, and the safety analysis can be adapted and re-executed automatically when the system design (i.e.~the system model) is modified.

\paragraph{Realization/Implementation:}
In this step, the specified safety-relevant system is realized in hardware and software. For instance, the software is generated from the given behavioral models.

\paragraph{Test and Verification:}
Within the module test, the software integration test, and the system integration test phases, the developed system (hardware and software) is veriﬁed against the specification. For veriﬁcation techniques (such as source code veriﬁcation, unit testing, integration testing, etc.) various test automation techniques and frameworks are existing and can be used according to the recommendations in safety standards. In addition, testing of system safety mechanisms must automated. Since test data can be derived from the safety analysis models in form of CFTs, the required test data to ensure the correctness of the implemented safety mechanisms at system-level can be generated automatically \cite{Zeller2015a}. The tests can be executed as Software-in-the-Loop (SiL) tests using the generated code running in a simulation environment \cite{ReiterZHVBR17} or as Hardware-in-the-loop (HiL) tests where the test cases can be executed non-intrusively and side-effect free using the data probe concept \cite{froehlich16}.

\paragraph{Model-based Safety Argumentation:}
Finally, the safety-relevant system is validated, and a so-called \emph{Safety Case} is compiled to argue that the system is safe.
Based on the DDI concept, it is possible to automate the creation of the safety case.
Since safety requirements are provided as an input to the continuous safety assessment pipeline, the basic structure of the safety argumentation (i.e.~the claims derived form the safety requirements of the systems and regulatory rules) can be defined manually and only needs to be adopted if requirements are changing. Moreover, the evidences to prove that the claims are fulfilled are crated automatically in the pipeline (in the safety analysis and test \& verification steps). According to \cite{Reich2019} the relationships between the claims in the safety argumentation and evidences provided by the artifacts represented in an ODE model can created automatically. The resulting safety case can then be checked w.r.t.~consistency and completeness \cite{foster2020integration,9307709}. The DDI concept also eases the effort in case of modifications, since the impacts of changes in the safety argumentation or the evidences can be identified automatically due to the traceability established by the DDI \cite{ReichFCZR20}.

\paragraph{Assessment/Certification:}
The entire artifacts, processes, and tools of the system development life-cycle are subject to independent reviews and independent assessments. The documents necessary for the assessment or certification process (at least the technical part) can be generated from the model-based safety case created by the continuous safety assessment pipeline. Hence, assessors can be easily integrated in the development process and can receive the latest information in a well established form at the end of the sprint or at dedicated points in time during the development without additional overhead for the developers to create the necessary documentation.

\section{Towards DevOps for Safety-critical Systems}
\label{sec:SafeDevOps}
Although contineous safety assessment is the first steps towards DevOps in the context of safety-relevant systems, there are additional challenges which must be solved: 

\paragraph{Culture change in assessment \& certification organizations:}
To fully leverage the benefits of DevOps in the context of safety-relevant systems, the certification process must be integrated into the continuous delivery process.
Therefore, the culture of a so-called \emph{"big-bang"} all-at-once certification before the release \cite{steghoefer2019} needs to be changed and assessors should review the status of the project in terms of functional safety frequently. Hence, in context of DevOps, there is a need to integrate assessor as stakeholders into the contineous delivery process. In \cite{hanssen2018safescrum,6606635}, it is recommended that the assessor is incorporated into the agile development process, such that they are able to review the status of the project in terms of functional safety frequently. Therefore, we propose to explicitly add the role of the "Independent Safety Assessor" in a framework such as the Scaled Agility Framework (SAFe)\footnote{https://www.scaledagileframework.com/} . Hence, providing the independent assessor the possibility to continuously review documents or models and enabling transparency of the sprint and release planning. Thereby, it is possible to (1) make the assessor aware of ongoing activities and (2) ensure that the assessor can plan appropriate time slots for the review of the safety case and the approval or formal certification of the release.

\paragraph{Automated Change Impact Analysis (CIA):}
Since safety standards require a thorough Change Impact Analysis (CIA) for every change, automating CIA is essential to realize continuous delivery in safety-relevant systems.
To automatically perform a CIA, all the steps of the safety engineering life-cycle (e.g.~HARA, safety analyses, etc.) which need to be updated must be identified. Afterwards, the required updates must be applied to the system. 

To automate CIA, we must enable strict traceability between the various artifacts of the system, software, and safety engineering life-cycle to determine the elements with the artifacts that are affected by change (e.g.~in the source code or in a requirements document). This requires a holistic meta-model to capture all the artifacts and their relationships created during a system's development life-cycle. Moreover, tool support must be provided to architects and developers for storing the information without any overhead \cite{Ratiu2015}. The concept of the DDI provides a first version of a holistic meta-model for providing strict traceability and enabling an automated CIA \cite{ReichFCZR20}.

\paragraph{Qualification of the tool chain:}
In order to be able to develop a safety-relevant system and to deliver a sound safety case in conjunction with each new release of the product, also the environment used to create the project requires specific attention. Hence, tool qualification is mandatory if the results of a tool are not reviewed by humans (e.g.~a compiler).
However, the qualification of tools is domain, company, business unit and even project specific. Domain specific check lists or tool qualification kits already exist, but a generalization of these approaches is currently not possible. 
Therefore, we propose to add a dedicated role of the "Certification Engineer/Manager" in a framework such as the Scaled Agility Framework (SAFe). The Certification Engineer must responsible for establishing and maintaining the regulatory compliance of the development environment (including the required documentation) within the agile development project.

\section{Conclusions and Future Work}
\label{sec:summary}
Safety-critical systems must meet regulatory requirements and shall comply to safety standards. Moreover, the continuous delivery process of a new software version must be lifted from software to system level and people from different engineering disciplines must be involved.
Thus, applying agile development and the concepts of continuous delivery in context of safety-critical, software-intensive systems requires to solve specific challenges. In this paper, we summarize the challenges which need to be solved in order to realize continuous delivery of software in safety-critical systems. 
Moreover, we sketch potential approaches how to overcome the mentioned challenges based on existing model-based safety analysis approaches (such as Component Fault Trees) and model-based assurance cases in form of Digital Dependability Identities (DDIs). Thus, we are able to create a pipeline for continuous safety assessment for safety-critical software-intensive systems. 

Furthermore, we discuss additional steps towards DevOps for safety-critical systems beyond the automation of the safety assessment process. This includes the mandatory qualification of the tools since delivery pipeline itself becomes a subject to regulatory compliance, the integration of assessors as stakeholders into the continuous delivery process, and the automation of change impact analysis for changes of the software after the system is released (feedback for Ops). These topics will be addressed in future research as well as the integration of the continuous safety assessment pipeline in the Scaled Agility Framework (SAFe). Moreover, we will investigate the extension of Machine Learning Operations (MLOps) w.r.t.~safety-critical environments. 

%
%
%
\bibliographystyle{splncs04}
\bibliography{references}
\end{document}